%% file: epp.tex
\documentclass[nofootinbib,hyperref,letterpaper,floatfix,reprint,aps,pre,superscriptaddress]{revtex4-1}

\usepackage{graphicx}
\usepackage{amssymb,amsmath,amsfonts,gensymb}
\usepackage{hyperref}
\usepackage{color}
\usepackage[normalem]{ulem}
\usepackage{tikz}
\usetikzlibrary{arrows,shapes,positioning,calc,3d}
\usepackage{makecell}
\usepackage{array}
\newcolumntype{M}{>{$\vcenter\bgroup\hbox\bgroup}c<{\egroup\egroup$}}

\newcommand{\eg}{\textit{e.g.}}

\newcommand{\etal}{\textit{et al.}}
\newcommand{\etc}{\textit{etc.}}
\newcommand{\cn}[1]{\scalebox{0.01}{\hphantom{\cite{#1}}}[\kern-0.4em\citenum{#1}]}
\newcommand{\fref}[1]{Fig.\ \ref{#1}}

\newcommand{\eref}[1]{Eq.\ \eqref{#1}}

\definecolor{gcb1}{HTML}{A6CEE3} 
\definecolor{gcb2}{HTML}{1F78B4} 
\definecolor{gcb3}{HTML}{B2DF8A} 
\definecolor{gcb4}{HTML}{33A02C} 
\definecolor{gcb5}{HTML}{FB9A99} 
\definecolor{gcb6}{HTML}{E31A1C} 
\definecolor{gcb7}{HTML}{6A3D9A} 
\definecolor{gcb8}{HTML}{FF7F00} 
\definecolor{gcb9}{HTML}{CAB2D6} 
\definecolor{gcb10}{HTML}{FDBF6F} 

\begin{document}
\preprint{arXiv:1304.7545}
\title{Entropically Patchy Particles: Engineering Valence Through Shape Entropy}
\author{Greg van Anders}
\affiliation{Department of Chemical Engineering,
University of Michigan, Ann Arbor, MI 48109-2136, USA}
\author{N.\ Khalid Ahmed}
\affiliation{Department of Chemical Engineering,
University of Michigan, Ann Arbor, MI 48109-2136, USA}
\author{Ross Smith}
\affiliation{Department of Materials Science and Engineering,
University of Michigan, Ann Arbor, MI 48109-2136, USA}
\author{Michael Engel}
\affiliation{Department of Chemical Engineering,
University of Michigan, Ann Arbor, MI 48109-2136, USA}
\author{Sharon C.\ Glotzer}
\email{sglotzer@umich.edu}
\affiliation{Department of Chemical Engineering,
University of Michigan, Ann Arbor, MI 48109-2136, USA}
\affiliation{Department of Materials Science and Engineering,
University of Michigan, Ann Arbor, MI 48109-2136, USA}
\begin{abstract}
 Patchy particles are a popular paradigm for the design and synthesis of
 nanoparticles and colloids for self-assembly. In ``traditional'' patchy
 particles, anisotropic interactions arising from patterned coatings,
 functionalized molecules, DNA, and other enthalpic means create the possibility
 for directional binding of particles into higher-ordered structures. Although
 the anisotropic geometry of non-spherical particles contributes to the
 interaction patchiness through van der Waals, electrostatic, and other
 interactions, how particle shape contributes entropically to self-assembly is
 only now beginning to be understood. It has been recently demonstrated that,
 for hard shapes, entropic forces are directional. A newly proposed theoretical
 framework that defines and quantifies directional entropic forces demonstrates
 the anisotropic--that is, patchy--nature of these emergent, attractive forces.
 Here we introduce the notion of entropically patchy particles as the entropic
 counterpart to enthalpically patchy particles. Using three example ``families''
 of shapes, we judiciously modify entropic patchiness by introducing geometric
 features to the particles so as to target specific crystal structures, which
 then assembled with Monte Carlo simulations. We quantify the emergent entropic
 valence via a potential of mean force and torque.  We generalize these shape
 operations to shape anisotropy dimensions, in analogy with the anisotropy
 dimensions introduced for enthalpically patchy particles.  Our findings
 demonstrate that entropic patchiness and emergent valence provide a way of
 engineering directional bonding into nanoparticle systems, whether in the
 presence or absence of additional, non-entropic forces. 
\end{abstract}
\maketitle

\begin{figure}
  \begin{center}
    \includegraphics[width=9cm]{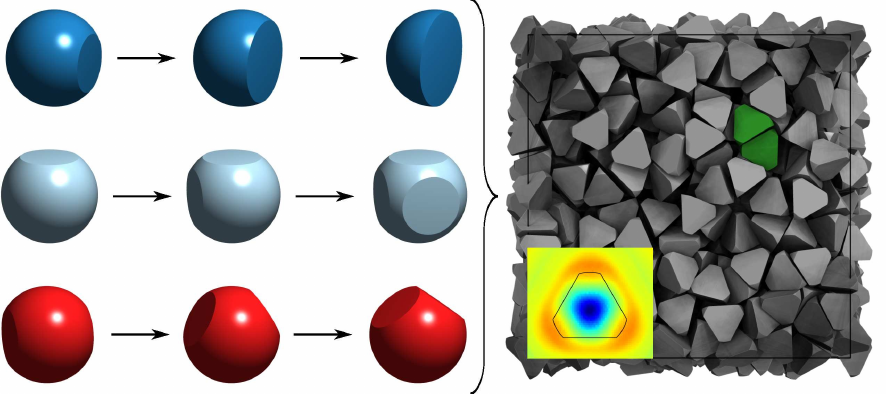}
  \end{center}
\end{figure}
\section{Introduction}

Patchy particles\cite{patchy,glotzsolomon} self assemble into nanoparticle
superlattices and colloidal crystals by exploiting anisotropic interactions
arising from, \eg, molecular patterning, DNA functionalization, and charge heterogeneity. Examples of patchy particles include Janus colloids
\cite{cayrejanus,cayreetal,Roh2005,zhangetal1,cui2006,granickjanussynth,
januscube},
striped nanospheres \cite{jackson04}
and nanorods \cite{mallouk},
and DNA-coated patchy particles \cite{pinedna},
among many others\cite{glotzsolomon,pawarkretzchmar}.
When the particles are not spheres,
non-isotropic van der Waals and other forces also contribute to interaction
patchiness.  With patchy particles, anisotropically placed patches promoting
either specific or nonspecific interactions with patches on other particles
induce directional ``bonding'' between particles of the sort typically
attributed to molecular substances. To date, patchy particles have been
assembled into numerous structures
\cite{mirkinmetalpoly,steinjacs,microchain,endtoend,agcubefunc,zhangetal2,velev,
granickkagome,granickjanusscience,wangrev,steinrev},
many of them isostructural to their atomic and molecular counterparts. 

Recently, there has been considerable focus on the contribution of entropic
forces to the assembly of anisotropically shaped particles into complex
structures\cite{freons,strlekfresph,schillingpent,pped,amirnature,dijkstraentdr,
escobedo,
zhaosquare,amirprl,rossi,trunctet,zoopaper,geissleryang,dijkstrasuperballs,nanooct,escopoly,dijkstracube,dijkstraplatelet,dijkstratcube,shape4sa}.
A general observation from many of these studies is that dense suspensions of
hard, faceted particles align their facets so as to maximize the system entropy,
giving rise to ordered structures as complex as colloidal quasicrystals
\cite{amirnature,trunctet} and crystals with unit cells containing as many as 52
particles \cite{zoopaper}. Damasceno, \etal\ \cite{trunctet,zoopaper}
rationalized this tendency toward facet alignment as the emergence of
``directional entropic forces'' between hard particles. Directional entropic
forces (DEFs) are not intrinsic to the particles, but instead are statistical
and emerge from the collective behavior of the entire system upon crowding. 
The DEF approach to the self-assembly of colloidal cubes, octahedra, rhombic
dodecahedra, and tetrahexahedra was recently demonstrated by Young et al.
\cite{kayliedef}.

A theoretical framework for the description and quantification of DEFs was
proposed recently\cite{entint}.  Using an effective potential of mean force
and torque (PMFT), hard polyhedra, spherocylinders, and hemispheres were shown
to exhibit spatially anisotropic probability distributions describing the likely
positions of neighboring particles, much like the polyvalent nature of
molecules. Unlike in molecular systems, however, here the valence is density
dependent and--because it arises statistically from collective behavior--is
emergent. We define these emergent valence regions of effective attraction
between particles as ``entropic patches''. These patches can achieve strengths
of several to many $k_\text{B}T$, aligning complementary geometrical features
just as enthalpically patchy particles align complementary enthalpic features.   

Here we introduce the notion of entropically patchy particles as the entropic
counterpart to enthalpically patchy particles, and we show how DEFs can be
engineered through the systematic alteration of particle shape to target
specific self-assembled structures. Using three example families of shapes, we
systematically apply certain shape operations to the particles so as modify
entropic patchiness to be consistent with target specific crystal structures. In
contrast to the charge, chemical, \etc\ mediation of the interaction between
sticky patches on \emph{enthalpically} patchy particles, attractive
\emph{entropic} patches are features in particle shape that promote local dense
packing in thermodynamic equilibrium. We show that we indeed obtain these
structures through self assembly with Monte Carlo (MC) simulations. We
generalize these shape operations to shape anisotropy dimensions, in analogy
with the anisotropy dimensions introduced for patchy
particles\cite{glotzsolomon}. Our findings demonstrate the utility of the
notions of entropic patchiness and emergent valence as an additional way of
engineering directional bonding into nanoparticle systems, whether in the
presence or absence of additional, non-entropic forces.

\section{Background}
The PMFT describing the directional entropic force between a pair of hard
particles in a system of identical particles has been derived\cite{entint} as
\begin{equation} \label{FMFT}
    F_{12}(\Delta \xi_{12}) = 
    -k_\text{B}T\log(H(d(\Delta \xi_{12}))J(\Delta \xi_{12}))+
    \tilde F_{12}(\Delta \xi_{12}) \; .
\end{equation}
Here, $\Delta \xi_{12}$ describes the relative position and orientation of the
particle pair.  $H$ is the Heaviside step function, and $d(\Delta \xi_{12})$ is
the minimum separation distance of the particle pair in their relative position
and orientation, which is negative when the particles overlap, and positive when
they do not. $J$ is the Jacobian for the set of invariant coordinates of
interest for a particular problem. $\tilde F_{12}$ is the Helmholtz free energy
available to other particles in the system when the relative position and
orientation of the pair is fixed.

From \eref{FMFT}, the PMFT can be seen to arise as a competition between two
terms. The term coming from $\tilde F_{12}$ is determined by the free energy of
the system with the pair fixed. If there is no intrinsic attraction among the
particles, then this term will tend to induce a locally denser packing for the
particle pair. The other contribution comes from the preference of the pair
itself for a particular relative position and orientation.  $\tilde F_{12}$ will
be minimized when the particle pair aligns itself to maximize its local packing
density according to the shape of the particles. The features of particle
shape that facilitate locally dense packing, therefore, act as the ``source'' of
the emergent attractive DEFs. We refer to these features as entropic patches.

\section{Results and Discussion}
\subsection{Targeted Self-Assembly Through Emergent Valence}

We design entropic patches to self assemble the following target structures:
simple cubic, body-centered cubic, diamond, and dodecagonal quasicrystal.  Each
of these has been reported in experiments or simulations of patchy spheres or
hard polyhedra. In all cases, the local coordination shell at least partially
dictates the type of crystal structure that assembles \cite{zoopaper}. We
attempt to create similar local coordination shells through entropic patches
engineered by slicing facets into hard spheres. Specifically, we simulate
spheres with cubic, octahedral, and tetrahedral faceting to induce the
appropriate entropic patchiness, and show that at sufficient crowding the
desired valence emerges leading to crystallization of target structures
consistent with that particular polyvalent coordination via the organization of
successive neighbor shells. 

\begin{figure*}
  \begin{tabular}{ccc}
    \resizebox{2in}{!}{\input{tworendtf}} &
    \resizebox{2in}{!}{\input{tetw}} &
    \includegraphics[width=2in]{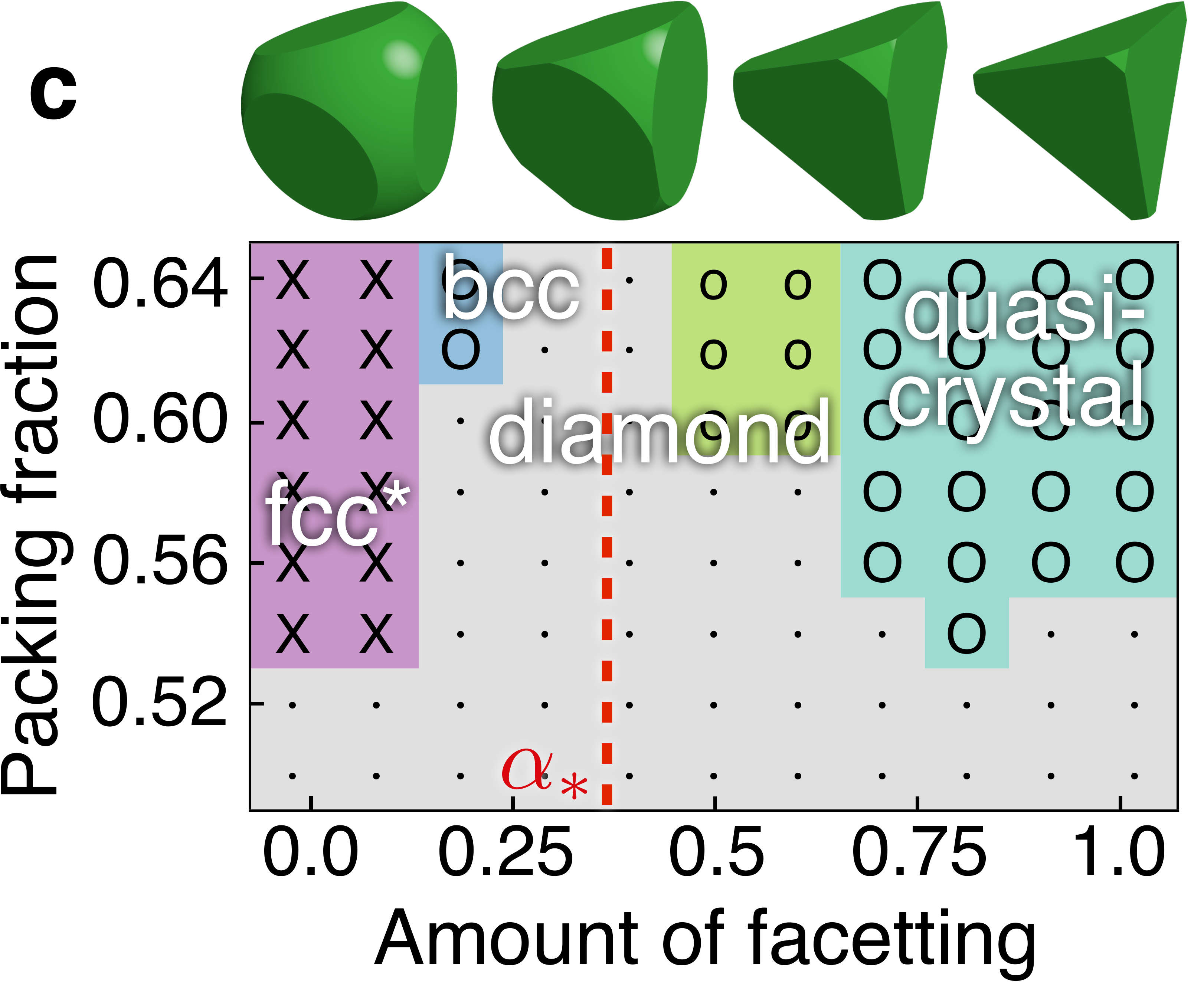} \\
    \resizebox{2in}{!}{\input{tfmap_xy_0}} &
    \resizebox{2in}{!}{\input{tfmap_xy_1}} &
    \resizebox{2in}{!}{\input{tfmap_xy_2}} \\
    \resizebox{2in}{!}{\input{tfmap_yz_0}} &
    \resizebox{2in}{!}{\input{tfmap_yz_1}} &
    \resizebox{2in}{!}{\input{tfmap_yz_2}} \\
  \end{tabular}
  \caption{We demonstrate shape-induced entropic valence in monodisperse
  systems of hard
  tetrahedrally
  faceted spheres by computing the PMFT for a pair of particles in a system of
  1000 particles at packing fraction 0.5 (a).
  Slicing the potential along two different planes (b) shows the induced valence
  in a plane through the facet (d, e, and f) and a plane parallel to the facet
  (g, h, and i). At sufficient crowding, the entropic forces arising from shape
  entropy lead to crystallization (c).
  As the faceting amount increases from
  $0.1$ (d and g),
  to
  $0.6$ (e and h),
  to
  $0.8$ (f and i),
  the PMFT shows greater evidence of shape-induced entropic valence that
  determines the
  crystal structure, even at insufficient crowding ($50\%$) to provoke
  crystallization.
  \label{patchstr-t}}
\end{figure*}
We first target the assembly of a (tetrahedrally-coordinated) diamond lattice by
tetrahedrally faceting spheres. The diamond lattice has been assembled in
simulation by decorating a sphere with four tetrahedrally coordinated enthalpic
patches (sticky spots) \cite{patchy,romanosciortino}. We therefore ``slice''
four equal sized facets into a sphere at the locations of the faces of a regular
tetrahedron.  We consider a faceting amount of $0$ to be a perfect sphere, and
$1$ to be a perfect tetrahedron. For concreteness, consider a tetrahedron with
vertices at $(1,1,1)$, $(-1,-1,1)$, $(-1,1,-1)$, and $(1,-1,-1)$. A perfect
sphere is the intersection of this tetrahedron with a sphere centered about the
origin with radius $\tfrac{1}{\sqrt{3}}$. A perfect tetrahedron is the
intersection of this tetrahedron and a sphere with radius $\sqrt{3}$. The radius
of the sphere required to generate any amount of faceting $\alpha$ between these
limits is given by the formula $\tfrac{1}{\sqrt{3}}(1-\alpha)+\sqrt{3}\alpha$.
We performed MC simulations of monodisperse systems of 1000 such particles at
fixed volume, for several choices of $\alpha$ (see Methods for details). We
computed the force component of the PMFT as a function of the Cartesian
components of the separation vector of the particles in the frame of one of the
particles. (A detailed description of the computation and a discussion of
possible coordinate systems have been given elsewhere\cite{entint}.) In
\fref{patchstr-t} we show that monodisperse, tetrahedrally faceted spheres
manifest shape-induced entropic valence (via the PMFT) in dense fluids at $50\%$
packing fraction.  Simulations were performed at faceting amounts between a
perfect sphere ($\alpha=0$) and a perfect tetrahedron ($\alpha=1$).\footnote{See
Table 1\ in Supporting Information for actual faceting amounts for this and
other particles below.} At a faceting amount of $0.6$ the particles
self-assemble a diamond lattice in MC simulation at a packing fraction of $60\%$
as shown in \fref{patchstr-t}c. Also note that when the faceting amount $\alpha$
exceeds
\begin{equation}
  \alpha_* = \frac{\sqrt{3}-1}{2} \approx 0.3660254 \, ,
\end{equation}
(dashed red line in \fref{patchstr-t}c) the faceting patches share adjacent
edges, which we would expect to have an effect on the local dense packing.  Only
above this faceting amount are we able to assemble the diamond lattice as
shown in \fref{patchstr-t}c. Note that a family of moderately truncated
tetrahedra also assemble a diamond lattice\cite{trunctet}.

\begin{figure*}
  \center
  \begin{tabular}{ccc}
    \resizebox{2in}{!}{\input{tworend}} &
    \resizebox{2in}{!}{\input{cubew}} &
    \includegraphics[width=2in]{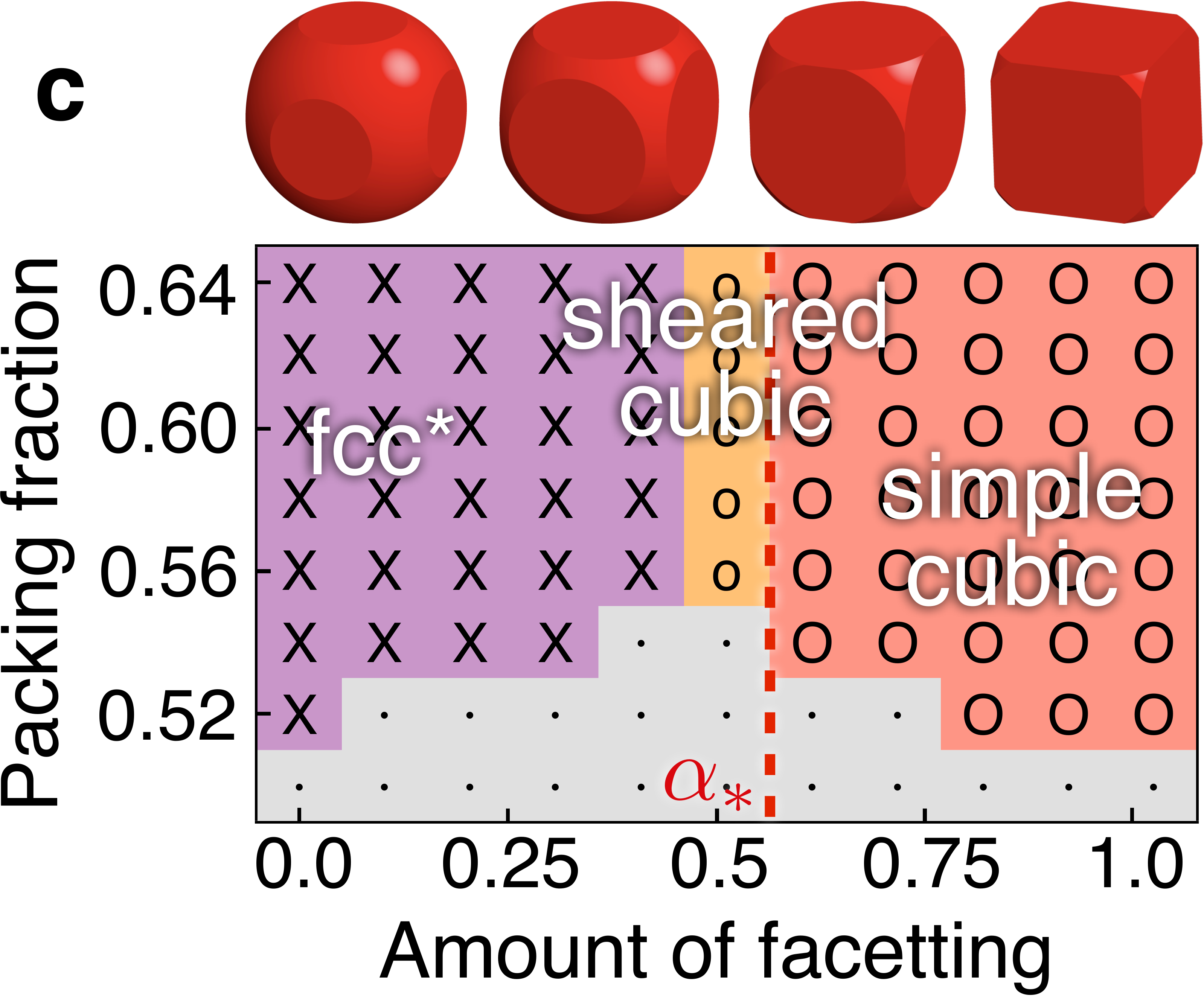}\\
    \resizebox{2in}{!}{\input{map_xy_0}} &
    \resizebox{2in}{!}{\input{map_xy_1}} &
    \resizebox{2in}{!}{\input{map_xy_2}} \\
    \resizebox{2in}{!}{\input{map_yz_0}} &
    \resizebox{2in}{!}{\input{map_yz_1}} &
    \resizebox{2in}{!}{\input{map_yz_2}} \\
  \end{tabular}
  \caption{We demonstrate shape-induced entropic valence in monodisperse
  systems of hard
  cubically
  faceted spheres by computing the PMFT for a pair of particles in a system of
  1000 particles at packing fraction 0.5 (a).
  Slicing the potential along two different planes (b) shows the induced valence
  in a plane through the facet (d, e, and f) and a plane parallel to the facet
  (g, h, and i). At sufficient crowding, the entropic forces arising from shape
  entropy lead to crystallization (c).
  As the faceting amount increases from
  $0.2$ (d and g),
  to
  $0.5$ (e and h),
  to
  $0.8$ (f and i),
  the PMFT shows greater evidence of shape-induced entropic valence that
  determines the
  crystal structure, even at insufficient crowding ($50\%$) to provoke
  crystallization.
  \label{patchstr-c}}
\end{figure*}
As a second example, we next target simple cubic lattices by slicing cubically
coordinated facets into hard spheres.  We denote a perfect sphere as $0$ and a
perfect cube as $1$. For concreteness, take the vertices of the cube
to be at $(1,1,1)$, and at similar locations in each of the other octants.
A perfect sphere is the intersection of this cube with a sphere centered about
the origin with unit radius. A perfect cube would be the intersection with a
sphere of radius $\sqrt{3}$. The radius of the sphere required to generate any
amount of faceting $\alpha$ in between a cube and a sphere is given by the
formula $1+(\sqrt{3}-1)\alpha$.  In \fref{patchstr-c} we show the role of the
PMFT in generating directional entropic forces that cause particles to have high
positional correlation at facet locations.

We compute the PMFT in a monodisperse system of hard cubically faceted spheres
at a density of $50\%$ (in the fluid phase) at faceting amounts of $0.2$, $0.5$,
and $0.8$.\footnote{See supporting information for other faceting amounts.} We
plot the PMFT in panels d-i, by slicing according to the diagram in
\fref{patchstr-c}b with faceting amount $\alpha$ increasing from left to right.
In panels d-f (label ed also with blue), we show slices of the PMFT at constant
$z$ as depicted by the blue plane in \fref{patchstr-c}b.  (Here we take
$z\approx 0$.) In panels g-i (labelled also with mauve) we slice the PMFT
parallel to the faceting patch (the mauve plane in \fref{patchstr-c}b) through
the minimum of the potential.  We see that upon increasing the faceting amount
we induce a greater amount of cubic coordination in the fluid at fixed density.
At a faceting amount of $0.2$ (panels d and g), the PMFT is nearly isotropic,
indicating that faceting plays only a small role in locally ordering the
particles.  However, at faceting amounts of $0.5$ (panels e and h) and $0.8$
(panels f and i) we observe PMFT differences on the order of $2-3\, k_\text{B}T$
favoring alignment of facets.  Simple cubic lattices (\fref{patchstr-c}c)
assemble at $\alpha \approx 0.6$ or more in MC simulations at packing fractions
of $54\%$ or more. Note that when the faceting amount $\alpha$ exceeds
\begin{equation} \alpha_* = \frac{\sqrt{2}-1}{\sqrt{3}-1} \approx 0.565826 \, ,
\end{equation} (dashed red line in \fref{patchstr-c}c) the shape of the faceting
patch goes from being circular to having four straight edges that are shared by
adjacent patches. The existence of this edge should have an effect on the
geometry of the locally preferred packing, and it is, perhaps, not surprising
that above this faceting amount we observe the assembly of simple cubic
lattices. The self-assembly of cubic, or nearly cubic particles into simple
cubic lattices has been seen before\cite{dijkstrasuperballs,dijkstracube}.

\begin{figure*}
  \begin{tabular}{ccc}
    \resizebox{2in}{!}{\input{tworendof}} &
    \scalebox{0.85}{\input{octw}} &
    \includegraphics[width=2in]{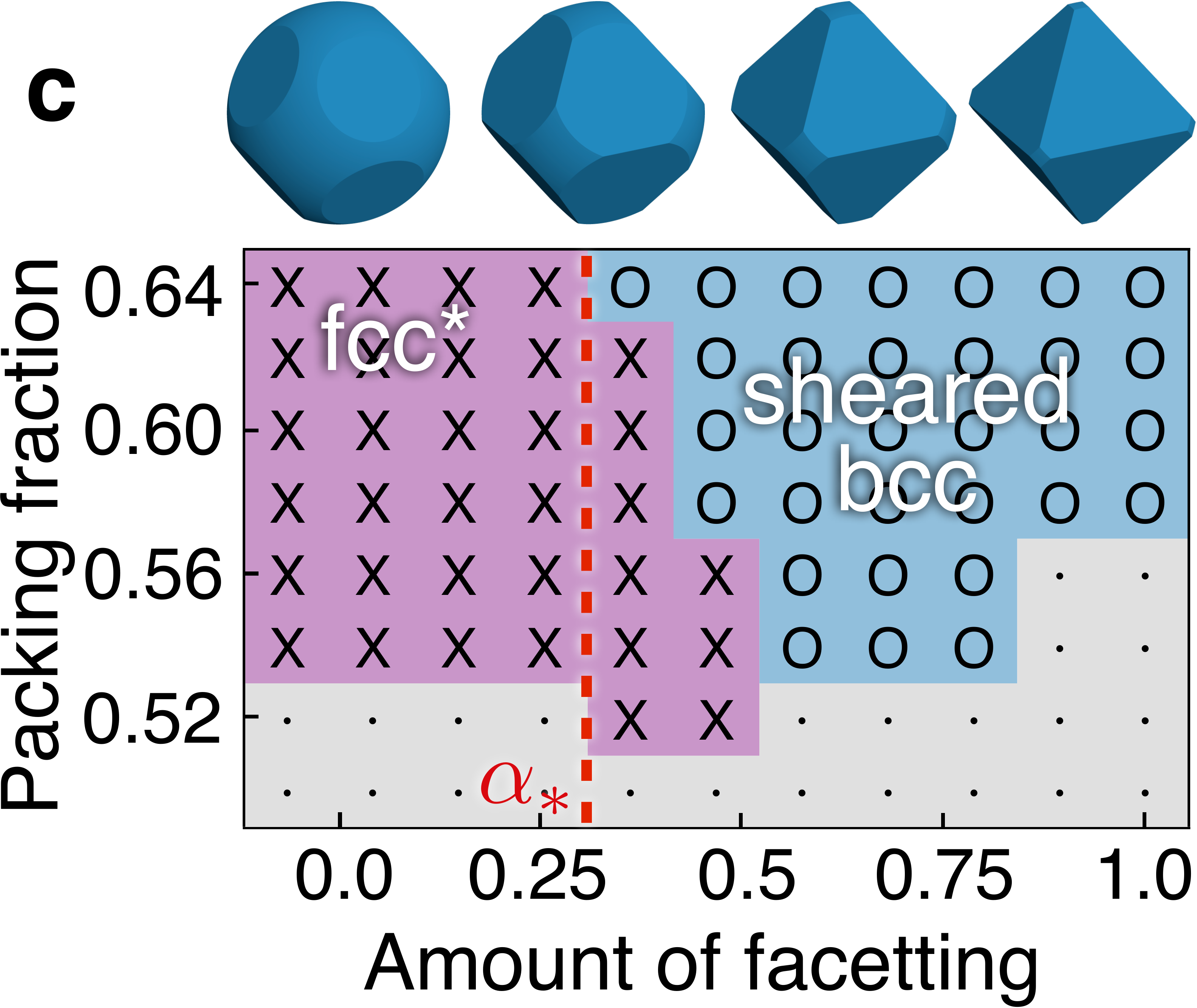} \\
    \resizebox{2in}{!}{\input{ofmap_xy_0}} &
    \resizebox{2in}{!}{\input{ofmap_xy_1}} &
    \resizebox{2in}{!}{\input{ofmap_xy_2}} \\
    \resizebox{2in}{!}{\input{ofmap_yz_0}} &
    \resizebox{2in}{!}{\input{ofmap_yz_1}} &
    \resizebox{2in}{!}{\input{ofmap_yz_2}} \\
  \end{tabular}
  \caption{We demonstrate shape-induced entropic valence in monodisperse
  systems of hard
  octahedrally
  faceted spheres by computing the PMFT for a pair of particles in a system of
  1000 particles at packing fraction 0.5 (a).
  Slicing the potential along two different planes (b) shows the induced valence
  in a plane through the facet (d, e, and f) and a plane parallel to the facet
  (g, h, and i). At sufficient crowding, the entropic forces arising from shape
  entropy lead to crystallization (c).
  As the faceting amount increases from
  $0.2$ (d and g),
  to
  $0.6$ (e and h),
  to
  $0.8$ (f and i),
  the PMFT shows greater evidence of shape-induced entropic valence that
  determines the
  crystal structure, even at insufficient crowding ($50\%$) to provoke
  crystallization.
  \label{patchstr-o}}
\end{figure*}
As a final example, we target a bcc or sheared bcc crystal with eight nearest
neighbors by octahedrally faceting a sphere. We denote a perfect sphere as $0$
and a perfect octahedron as $1$. For concreteness, take the vertices to lie at a
unit distance from the origin along each of the Cartesian coordinate axes, at
$(1,0,0)$, $(-1,0,0)$, \etc\ A perfect sphere is the intersection of this cube
with a sphere centered about the origin with radius $\tfrac{1}{\sqrt{3}}$.  A
perfect octahedron is the intersection with a sphere of unit radius. The radius
of the sphere required to generate any amount of faceting $\alpha$ between the
sphere and the octahedron is given by the formula
$1+(1-\tfrac{1}{\sqrt{3}})\alpha$. Simulations were performed at faceting
amounts between these two limits. At low faceting amounts, where the entropic
valence (patchiness) is fairly isotropic, the systems assembled fcc crystals.
However, at faceting amounts as low as $0.4$, we observe sheared bcc crystals.
Note that when the faceting amount $\alpha$ exceeds
\begin{equation}
  \alpha_* = \frac{\frac{\sqrt{3}}{\sqrt{2}}-1}{\sqrt{3}-1}
  \approx 0.307007 \, ,
\end{equation}
(dashed red line in \fref{patchstr-o}c) the faceting patches begin to share
adjacent edges. This figure coincides with the lowest faceting amount at which
we observed the sheared bcc crystal in our simulations. Octahedra have been
studied previously,
\cite{trunctet,dijkstrasuperballs,geissleryang,zoopaper,dijkstratcube} where bcc
and sheared bcc lattices were also observed. One interesting feature of the
octahedrally faceted particles is that entropically preferred local dense
packings have particles situated face-to-face, but with the orientation of the
adjacent faces rotated by 180\degree. In the resulting ``star of David''
arrangement (see Fig.~S10) the protruding vertices reduce the free volume
available to the surrounding particles. This means that although face-to-face
arrangements are still favored by shape entropy, as exhibited in the increasing
anisotropy of the PMFT depicted in \fref{patchstr-o}, the strength of the
entropic patch is actually lessened at high degrees of faceting, compared with
lower degrees.

\subsection{Generalization and Anisotropy Dimensions}
\begin{figure*}
  \centering
  \scalebox{0.75}{\includegraphics{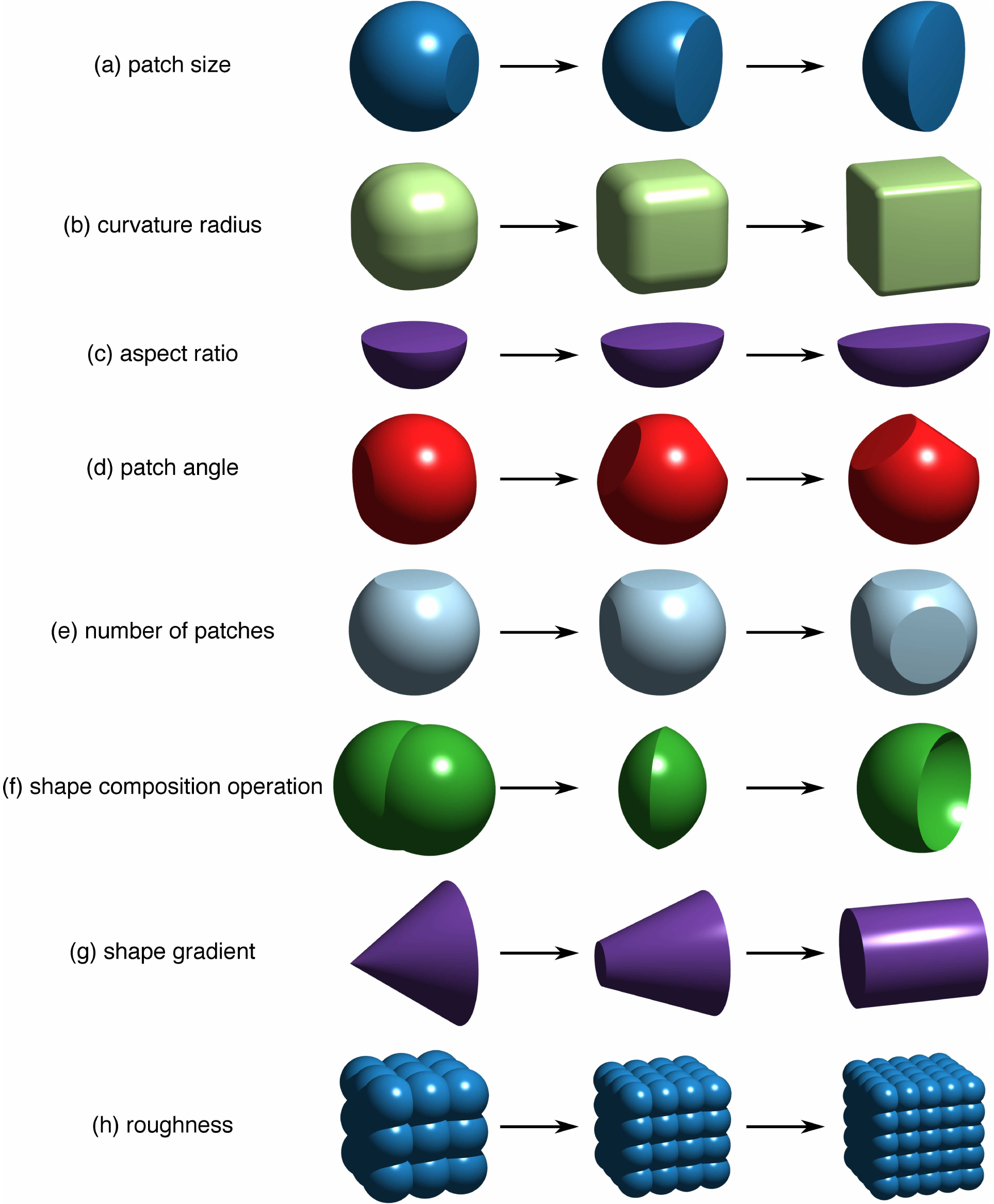}}
  \caption{(a) The creation and altering of entropic patches can be
  conceptualized as anisotropy dimensions:
  (a) patch size, (b) curvature radius, (c) aspect ratio, and (d) patch
  angle, (e) number of patches, (f) shape composition operation,
  (g) shape gradient, and (h) roughness.
  \label{anisodims}}
\end{figure*}
\begin{figure*}
  \includegraphics{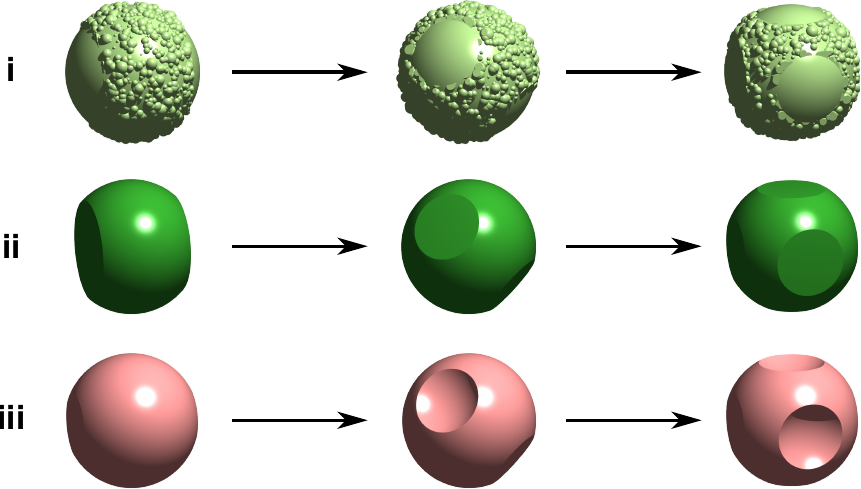}
  \caption{Schematic showing the the same anisotropy dimension (e, number of
  patches) applied to three different particle types: (i) a roughened colloid,
  (ii) a faceted sphere, and (iii) a dimpled sphere.
  \label{fig3}
  }
\end{figure*}
\begin{figure*}
  \includegraphics{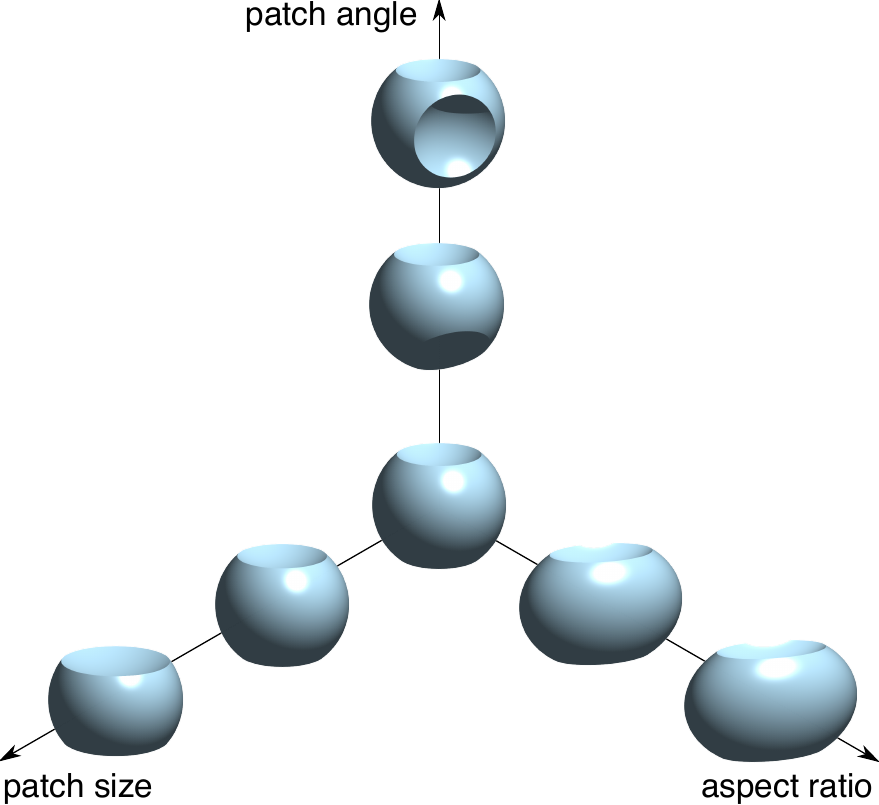}
  \caption{Schematic showing the result of combining various anisotropy
  dimensions by applying patch angle (d), aspect ratio (c), and patch size (a)
  from \fref{anisodims} to a sphere with two ``dimple'' patches.
  \label{fig4}
  }
\end{figure*}
\begin{figure}
  \centering
  \scalebox{0.85}{\includegraphics{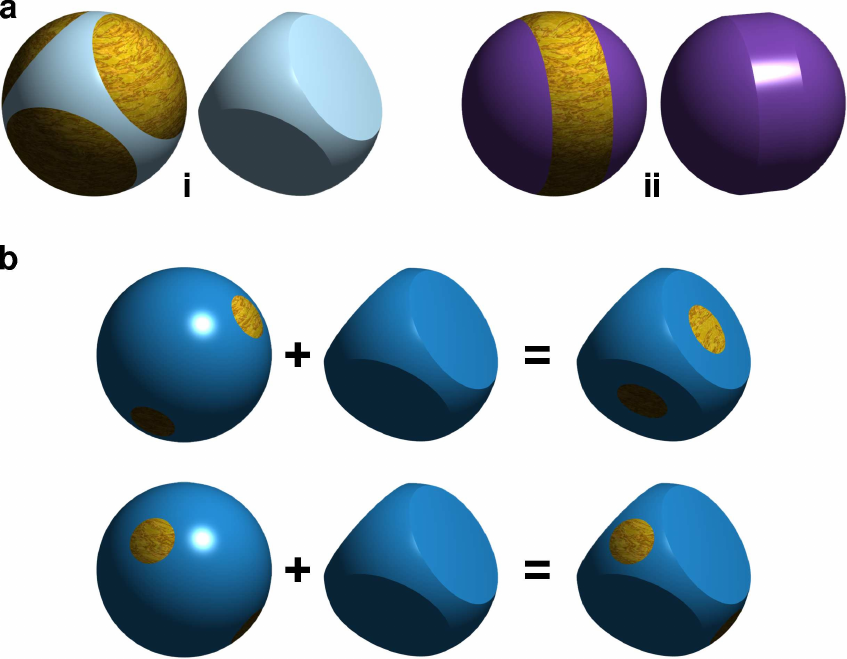}}
  \caption{Panel (a): Many anisotropy dimensions for enthalpically patchy
  particles \cite{glotzsolomon} have entropically patchy counterparts. Example
  traditional sticky patchy particles left (i) and (ii) with entropically patchy
  counterparts right (i) and (ii).  Panel (b): Enthalpic patches can be combined
  with entropic patches to enhance or inhibit entropic patchiness as shown
  schematically here. This can be obtained by using the same direction for both
  types of patches (top), or by using different directions for the entropic
  patches and the enthalpic patches (bottom).}
  \label{bothpatch}
\end{figure}
 
The angular specificity achieved by facet alignment via DEFs is reminiscent of
the angular specificity of enthalpic interactions conceptualized within the
patchy particle paradigm\cite{patchy,glotzsolomon}. However, in contrast to
the chemical or other patterning that leads to enthalpic patchiness, the angular
specificity of interactions arises here solely due to entropic considerations
arising from features in particle shape that promote local dense packing. These
features act as ``entropic patches'' that cause preferential alignment. Beyond
the simple faceting of hard spheres, there are many ways of altering particle
shape to introduce patches that promote local dense packing.

The various shape operations that may be applied to generate attractive entropic
patches may be described in terms of  anisotropy dimensions (\fref{anisodims}),
as was done for enthalpic patches \cite{glotzsolomon}. Of course, in contrast to
traditional sticky patches, entropic patchiness is an emergent, effective
concept that depends not only on density but also on all geometric features
within the characteristic length of the interaction. Such features may include,
\eg, flat facets or other low curvature regions, and interlocking or ``mating''
features. Conversely, high curvature regions could be used to introduce
repulsive patches.

Eight examples of shape anisotropy dimensions are illustrated in
\fref{anisodims}.  Many of these anisotropy dimensions have already been
explored in particles synthesized in the literature. For example, in
lock-and-key colloids \cite{pacman} the anisotropy dimensions of patch size (a),
curvature radius (b), and shape composition operation (f) have been synthesized
\cite{stefanosynth}.  In roughened colloids the anisotropy dimensions of patch
size (a), aspect ratio (c), patch angle (d), number of patches (e) and roughness
(h) have been synthesized
\cite{zhaomason1,zhaomason2,stroock1,stroock2,yakerough,synderrough,kraftetal}.
There are many other examples of work in the literature that can be considered
explorations of these anisotropy dimensions
\cite{ptnanoshape,saushapes,demortiere,auoct,auoctagcube,yangbinary,yangptnano,
skrabalaksynth,janaetal,yangplatonic,klajnetal,tianetal,murphyrev,
aushapecontrol,flamesynth,
hoellipsoid,whitesides2d,vbellipsoid1,vbellipsoid2,rodviz,colloidsoup,
hololitho,champion,nobumof,kayliedef,taoetal,shapeanisonanosa,goldbipyramid,
saurogach,skrabalak,multidimple}.

Shape anisotropy dimensions can be combined in different ways to yield various
particles that have appeared in the literature, \eg\ lock-and-key colloids
\cite{lockkeyent,pacman}, and various novel particle geometries.  For example,
in \fref{fig3} we consider applying the anisotropy dimension of number of
patches to three different species of particle: roughened colloids
\cite{zhaomason1,zhaomason2,stroock1,stroock2,yakerough,synderrough,kraftetal},
faceted spheres, and dimpled spheres \cite{pacman,stefanosynth,multidimple}.
In \fref{fig4} we consider the shape space defined
by the application of three orthogonal anisotropy dimensions to the dimpled sphere.  We also note that many of these anisotropy
dimensions map naturally to anisotropy dimensions introduced for enthalpically
patchy particles \cite{glotzsolomon}. Examples of enthalpically patchy particles
and their entropically patchy analogues are shown in \fref{bothpatch}a.  As depicted in \fref{bothpatch}b, combining both enthalpic and entropic patchiness provides opportunities for enhancing particular desired particle alignment, or produce competing forces that may give rise to structures of high complexity. 

As a final note, we point out that entropically patchy particles are also
relevant in systems with depletants, since depletant-induced colloidal
crystallization is controlled by the same entropic mechanism as the
crystallization of hard colloids.\cite{entint} The interaction range for DEFs is
determined by the scale of the particles being integrated out in the calculation
of the PMFT. In traditional depletion systems, the depletants are typically much
smaller than the colloids and penetrable. This hierarchy in scales produces a
very short ranged interaction between nanoparticles or colloids, which allows
for a clearer separation between geometric features on the colloids that
contribute to local dense packing, and thereby eases the identification of the
features that are the entropic patches. Particularly salient examples of
entropically patchy particles assembled through depletion forces include the
work on selectively roughened colloids
\cite{zhaomason1,zhaomason2,stroock1,stroock2,yakerough,synderrough,kraftetal}, lock-and-key colloids
\cite{lockkeyent,lockkey,pacman,pacsuprmol,stefanosynth,multidimple}, and
polyhedrally-shaped metal nanoparticles\cite{geissleryang,kayliedef}, each of
which exploits geometrical features to create anisotropy in entropic
interactions. These and related works have been reviewed elsewhere
\cite{shapecolloids}.

\section{Conclusion}
We showed in three example model systems that by judiciously engineering
particle shape we can induce angularly specific interactions between hard
particles strong enough and directional enough to induce the self-assembly of a
targeted crystal structure. The DEFs responsible for this ordering can be
measured experimentally using, \eg, optical tweezers and existing confocal
microscopy techniques \cite{intconf,rodviz}.  We abstracted these examples of
particle design to be part of a much broader method of shape engineering through
entropic patches that serves to provide a means of self-assembling materials
through entropic interactions alone.  We introduced anisotropy dimensions to
exhibit and organize the various ways in which particle shape can be engineered
to exploit this method of realizing complex structures.

When combined with traditional enthalpic patchiness, entropic patchiness greatly
enlarges the already vast design space for new nano- and micron-scale building
blocks. The theoretical framework based upon the PMFT\cite{entint}
allows the quantitative assessment of the relative strength of entropic driving
forces for assembly when other forces also contribute. At the nanoscale,
particles are seldom solely hard \cite{chs}.  However, the fact that DEFs cause
attractive interactions on the order of a few $k_\text{B} T$ at intermediate
packing densities that are easily experimentally accessible suggests there might
be a large class of systems in which electrostatic or other forces can be
sufficiently controlled such that entropic patches supply the dominant force
controlling their self-assembly, as was demonstrated recently
\cite{youngmirkin,glotzermurray,kayliedef}. When enthalpic interactions tend to
drive facet alignment anyway, as in, \eg, ligand-coated faceted nano particles
\cite{tang06-2,glotzermurray}, the entropic contribution will only enhance that
tendency. 

\section{Methods} \label{methods}
To measure the PMFT we performed MC simulations of dense fluids of $1000$ hard
faceted spheres at fixed volume. We note \cite{entint} that the PMFT can be
defined implicitly from the partition function
\begin{equation} \label{ZMFT}
  \mathcal{Z}
  = \int d(\Delta\xi_{12}) e^{-\beta F_{12}(\Delta\xi_{12})} \; .
\end{equation}
In this form the PMFT can be seen to be the logarithm of the integrand of the
partition function. We computed the PMFT by examining the displacement between
particles, which we computed in the coordinate frame of each particle, and
partitioned into defined regions. The orientations of the second particle were
integrated over so as to sufficiently reduce the dimensionality of the potential
to allow direct visualization. The possible relative positions of the particles
were subdivided into a number of regions, and the PMFT was computed by observing
the relative frequency of observing a pair of particles in each of these
regions, as indicated in \eref{ZMFT}.  Errors quoted are standard errors of the
mean of independent runs of independently equilibrated systems.  Full details on
the source and analysis of numerical errors that can arise in computing DEFs
can be found elsewhere \cite{entint}.

The MC method employed above, and for the results shown in
Figs.~\ref{patchstr-t}, \ref{patchstr-c}, and \ref{patchstr-o} employed single
particle moves for both translation and orientation. In all cases the simulation
box was taken to be periodic and the volume fixed. However, for
Figs.~\ref{patchstr-t}c, \ref{patchstr-c}c, and \ref{patchstr-o}c, the box was
permitted to shear at fixed volume.  Overlaps were checked using the same
implementation of the GJK algorithm \cite{gjk} used other work by some of the
present authors\cite{zoopaper}.

This document is the unedited Author's version of a Submitted Work that was
subsequently accepted for publication in ACS Nano, copyright (c) American
Chemical Society after peer review. To access the final edited and published
work see \href{http://dx.doi.org/10.1021/nn4057353}{DOI:10.1021/nn4057353}

\begin{acknowledgements}
We thank D.\ Klotsa and B.\ Schultz for helpful suggestions. This material is
based upon work supported by, or in part by, the U.S.\ Army Research Office
under Grant Award No.\ W911NF-10-1-0518, the DOD/ASD(R\&E) under Award No.\
N00244-09-1-0062, and the Biomolecular Materials Program of the Materials
Engineering and Science Division of Basic Energy Sciences at the U.S.\
Department of Energy under Grant No.\ DE-FG02-02ER46000. The calculations of
the hard particle PMFT were supported by ARO. The development of the PMFT and
anisotropy dimensions were supported by DoE. Any opinions, findings, and
conclusions or recommendations expressed in this publication are those of the
author(s) and do not necessarily reflect the views of the DOD/ASD(R\&E).
\end{acknowledgements}


%
\end{document}

%% file: tworendtf.tex
\begin{tikzpicture}
  \coordinate (o) at (0,0);
  \node[draw=none,fill=none,above right=0cm and 0.0cm of o]
  {\includegraphics[width=2in]{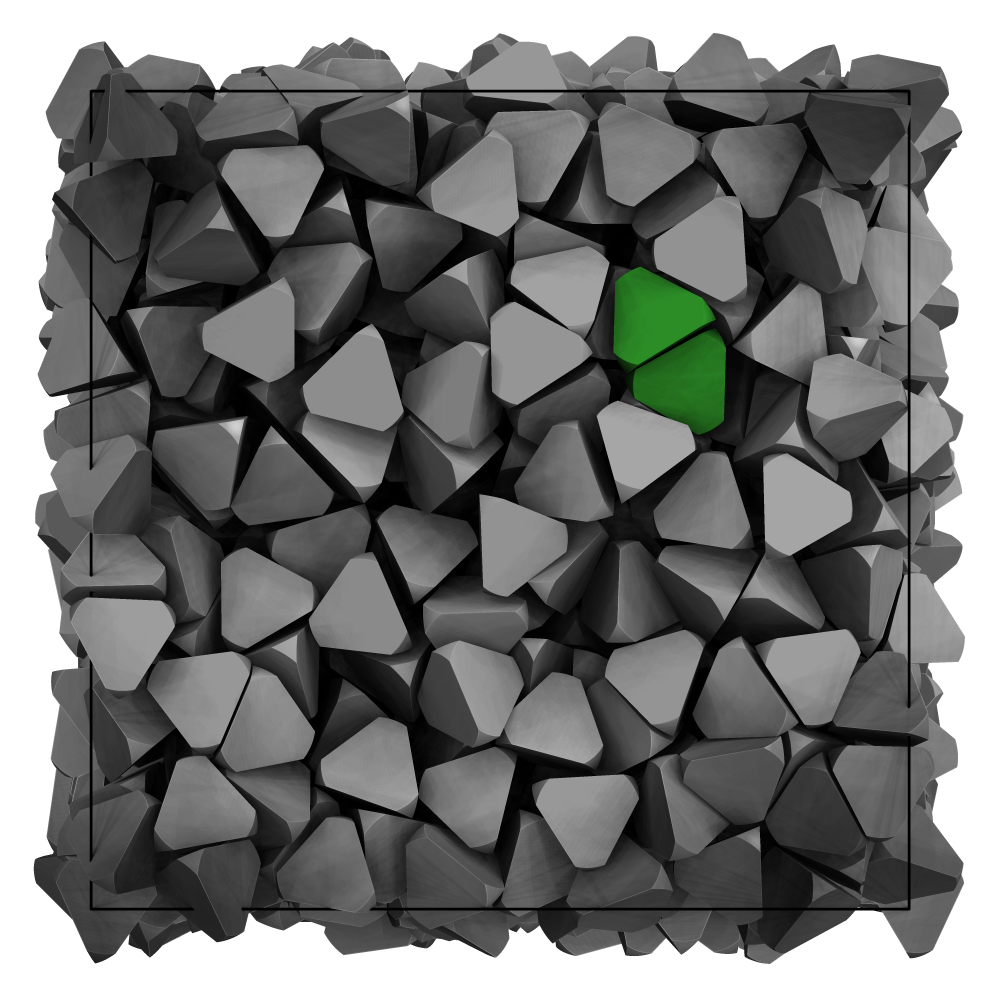}};
  \node[draw=none,fill=none,above right=1.75in and -0.2cm of o]
  {\textsf{\textbf{a}}};
\end{tikzpicture}

%% file: tetw.tex
\begin{tikzpicture}
  \coordinate (o) at (0,0);
  \node[draw=none,fill=none,above right=0cm and 0.0cm of o]
  {\makecell[c]{\scalebox{0.7}{\input{tet}}}};
  \node[draw=none,fill=none,above right=5.0cm and -0.2cm of o]
  {\textsf{\textbf{b}}};
\end{tikzpicture}

%% file: tfmap_xy_0.tex
\begin{tikzpicture}
  \coordinate (o) at (0,0);
  \node[draw=none,fill=none,above right=0cm and 0.0cm of o]
  {\includegraphics[width=2in]{{{tf2_xy_map_0.1-crop}}}};
  \node[draw=none,fill=none,above right=2.90cm and -0.1cm of o]
  {\includegraphics[width=0.22in]{{{tf_0.1-t}}}};
\end{tikzpicture}

%% file: tfmap_xy_1.tex
\begin{tikzpicture}
  \coordinate (o) at (0,0);
  \node[draw=none,fill=none,above right=0cm and 0.0cm of o]
  {\includegraphics[width=2in]{{{tf2_xy_map_0.6-crop}}}};
  \node[draw=none,fill=none,above right=2.90cm and -0.1cm of o]
  {\includegraphics[width=0.22in]{{{tf_0.6-t}}}};
\end{tikzpicture}

%% file: tfmap_xy_2.tex
\begin{tikzpicture}
  \coordinate (o) at (0,0);
  \node[draw=none,fill=none,above right=0cm and 0.0cm of o]
  {\includegraphics[width=2in]{{{tf2_xy_map_0.8-crop}}}};
  \node[draw=none,fill=none,above right=2.90cm and -0.1cm of o]
  {\includegraphics[width=0.22in]{{{tf_0.8-t}}}};
\end{tikzpicture}

%% file: tfmap_yz_0.tex
\begin{tikzpicture}
  \coordinate (o) at (0,0);
  \node[draw=none,fill=none,above right=0cm and 0.0cm of o]
  {\includegraphics[width=2in]{{{tf2map_0.1_1-crop}}}};
  \node[draw=none,fill=none,above right=2.70cm and -0.1cm of o]
  {\includegraphics[width=0.22in]{{{tf_0.1-t}}}};
\end{tikzpicture}

%% file: tfmap_yz_1.tex
\begin{tikzpicture}
  \coordinate (o) at (0,0);
  \node[draw=none,fill=none,above right=0cm and 0.0cm of o]
  {\includegraphics[width=2in]{{{tf2map_0.6_1-crop}}}};
  \node[draw=none,fill=none,above right=2.70cm and -0.1cm of o]
  {\includegraphics[width=0.22in]{{{tf_0.6-t}}}};
\end{tikzpicture}

%% file: tfmap_yz_2.tex
\begin{tikzpicture}
  \coordinate (o) at (0,0);
  \node[draw=none,fill=none,above right=0cm and 0.0cm of o]
  {\includegraphics[width=2in]{{{tf2map_0.8_1-crop}}}};
  \node[draw=none,fill=none,above right=2.70cm and -0.1cm of o]
  {\includegraphics[width=0.22in]{{{tf_0.8-t}}}};
\end{tikzpicture}

%% file: tworend.tex
\begin{tikzpicture}
  \coordinate (o) at (0,0);
  \node[draw=none,fill=none,above right=0cm and 0.0cm of o]
  {\includegraphics[width=5.5cm]{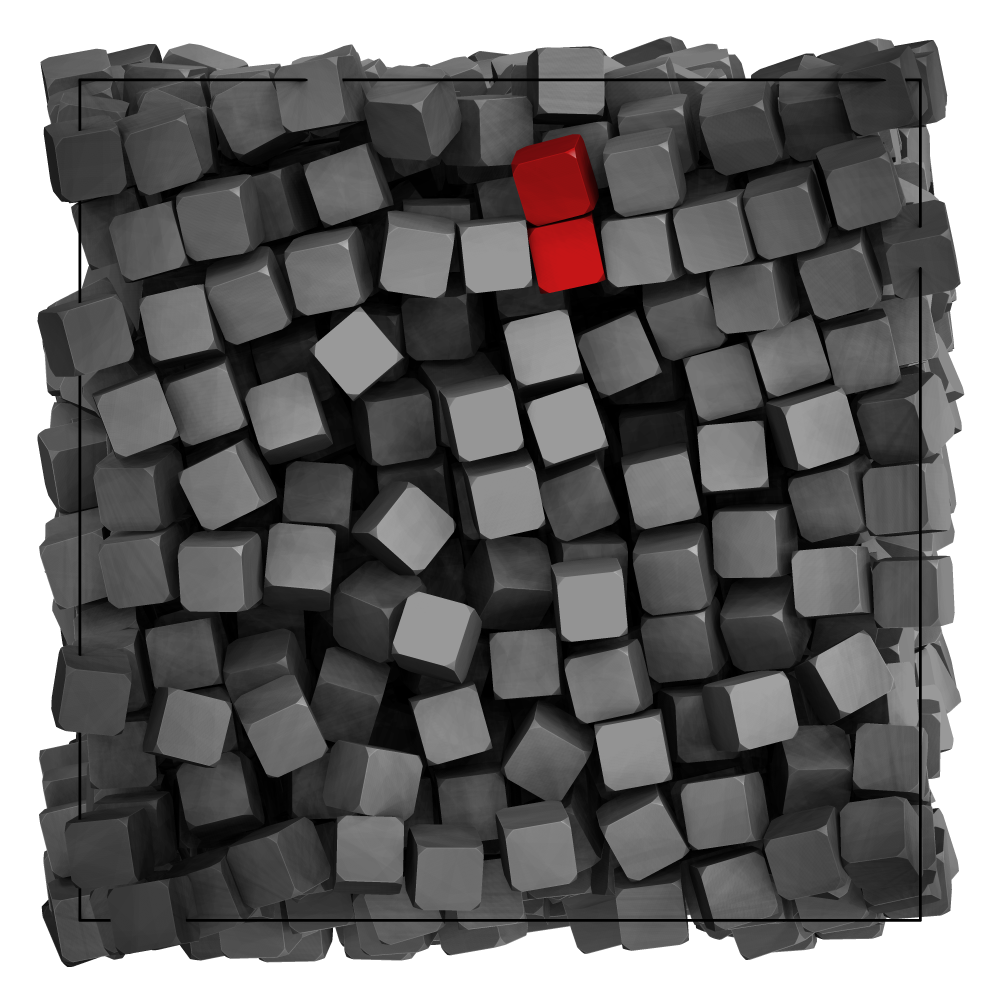}};
  \node[draw=none,fill=none,above right=5.0cm and -0.2cm of o]
  {\textsf{\textbf{a}}};
\end{tikzpicture}

%% file: cubew.tex
\begin{tikzpicture}
  \coordinate (o) at (0,0);
  \node[draw=none,fill=none,above right=0cm and 0.0cm of o]
  {\makecell[c]{\scalebox{0.7}{\input{cube}}}};
  \node[draw=none,fill=none,above right=5.0cm and -0.2cm of o]
  {\textsf{\textbf{b}}};
\end{tikzpicture}

%% file: map_xy_0.tex
\begin{tikzpicture}
  \coordinate (o) at (0,0);
  \node[draw=none,fill=none,above right=0cm and 0.0cm of o]
  {\includegraphics[width=2in]{{{cf_xy_map_0.2-crop}}}};
  \node[draw=none,fill=none,above right=2.75cm and -0.1cm of o]
  {\includegraphics[width=0.20in]{{{cf_0.2-t}}}};
\end{tikzpicture}

%% file: map_xy_1.tex
\begin{tikzpicture}
  \coordinate (o) at (0,0);
  \node[draw=none,fill=none,above right=0cm and 0.0cm of o]
  {\includegraphics[width=2in]{{{cf_xy_map_0.5-crop}}}};
  \node[draw=none,fill=none,above right=2.75cm and -0.1cm of o]
  {\includegraphics[width=0.20in]{{{cf_0.5-t}}}};
\end{tikzpicture}

%% file: map_xy_2.tex
\begin{tikzpicture}
  \coordinate (o) at (0,0);
  \node[draw=none,fill=none,above right=0cm and 0.0cm of o]
  {\includegraphics[width=2in]{{{cf_xy_map_0.8-crop}}}};
  \node[draw=none,fill=none,above right=2.75cm and -0.1cm of o]
  {\includegraphics[width=0.20in]{{{cf_0.8-t}}}};
\end{tikzpicture}

%% file: map_yz_0.tex
\begin{tikzpicture}
  \coordinate (o) at (0,0);
  \node[draw=none,fill=none,above right=0cm and 0.0cm of o]
  {\includegraphics[width=2in]{{{cf_yz_map_0.2-crop}}}};
  \node[draw=none,fill=none,above right=2.85cm and -0.1cm of o]
  {\includegraphics[width=0.20in]{{{cf_0.2-t}}}};
\end{tikzpicture}

%% file: map_yz_1.tex
\begin{tikzpicture}
  \coordinate (o) at (0,0);
  \node[draw=none,fill=none,above right=0cm and 0.0cm of o]
  {\includegraphics[width=2in]{{{cf_yz_map_0.5-crop}}}};
  \node[draw=none,fill=none,above right=2.85cm and -0.1cm of o]
  {\includegraphics[width=0.20in]{{{cf_0.5-t}}}};
\end{tikzpicture}

%% file: map_yz_2.tex
\begin{tikzpicture}
  \coordinate (o) at (0,0);
  \node[draw=none,fill=none,above right=0cm and 0.0cm of o]
  {\includegraphics[width=2in]{{{cf_yz_map_0.8-crop}}}};
  \node[draw=none,fill=none,above right=2.85cm and -0.1cm of o]
  {\includegraphics[width=0.20in]{{{cf_0.8-t}}}};
\end{tikzpicture}

%% file: tworendof.tex
\begin{tikzpicture}
  \coordinate (o) at (0,0);
  \node[draw=none,fill=none,above right=0cm and 0.0cm of o]
  {\includegraphics[width=2in]{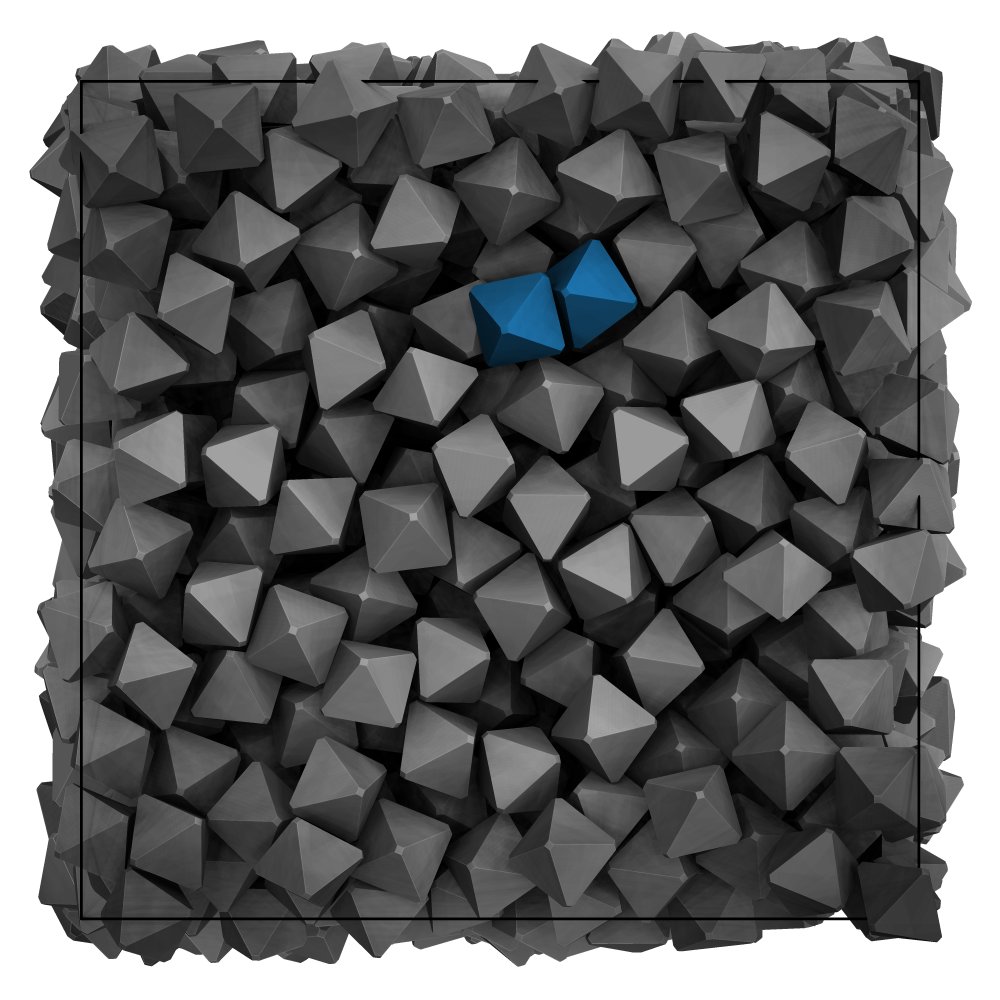}};
  \node[draw=none,fill=none,above right=1.75in and -0.2cm of o]
  {\textsf{\textbf{a}}};
\end{tikzpicture}

%% file: octw.tex
\begin{tikzpicture}
  \coordinate (o) at (0,0);
  \node[draw=none,fill=none,above right=0cm and 0.0cm of o]
  {\makecell[c]{\scalebox{0.7}{\input{oct}}}};
  \node[draw=none,fill=none,above right=3.5cm and -0.2cm of o]
  {\textsf{\textbf{b}}};
\end{tikzpicture}

%% file: ofmap_xy_0.tex
\begin{tikzpicture}
  \coordinate (o) at (0,0);
  \node[draw=none,fill=none,above right=0cm and 0.0cm of o]
  {\includegraphics[width=2in]{{{of_xy_map_0.2-crop}}}};
  \node[draw=none,fill=none,above right=2.70cm and -0.1cm of o]
  {\includegraphics[width=0.22in]{{{of_0.2-t}}}};
\end{tikzpicture}

%% file: ofmap_xy_1.tex
\begin{tikzpicture}
  \coordinate (o) at (0,0);
  \node[draw=none,fill=none,above right=0cm and 0.0cm of o]
  {\includegraphics[width=2in]{{{of_xy_map_0.5-crop}}}};
  \node[draw=none,fill=none,above right=2.70cm and -0.1cm of o]
  {\includegraphics[width=0.22in]{{{of_0.5-t}}}};
\end{tikzpicture}

%% file: ofmap_xy_2.tex
\begin{tikzpicture}
  \coordinate (o) at (0,0);
  \node[draw=none,fill=none,above right=0cm and 0.0cm of o]
  {\includegraphics[width=2in]{{{of_xy_map_0.8-crop}}}};
  \node[draw=none,fill=none,above right=2.70cm and -0.1cm of o]
  {\includegraphics[width=0.22in]{{{of_0.8-t}}}};
\end{tikzpicture}

%% file: ofmap_yz_0.tex
\begin{tikzpicture}
  \coordinate (o) at (0,0);
  \node[draw=none,fill=none,above right=0cm and 0.0cm of o]
  {\includegraphics[width=2in]{{{ofmap_0.2_1-crop}}}};
  \node[draw=none,fill=none,above right=2.80cm and -0.1cm of o]
  {\includegraphics[width=0.22in]{{{of_0.2-t}}}};
\end{tikzpicture}

%% file: ofmap_yz_1.tex
\begin{tikzpicture}
  \coordinate (o) at (0,0);
  \node[draw=none,fill=none,above right=0cm and 0.0cm of o]
  {\includegraphics[width=2in]{{{ofmap_0.5_2-crop}}}};
  \node[draw=none,fill=none,above right=2.80cm and -0.1cm of o]
  {\includegraphics[width=0.22in]{{{of_0.5-t}}}};
\end{tikzpicture}

%% file: ofmap_yz_2.tex
\begin{tikzpicture}
  \coordinate (o) at (0,0);
  \node[draw=none,fill=none,above right=0cm and 0.0cm of o]
  {\includegraphics[width=2in]{{{ofmap_0.8_2-crop}}}};
  \node[draw=none,fill=none,above right=2.80cm and -0.1cm of o]
  {\includegraphics[width=0.22in]{{{of_0.8-t}}}};
\end{tikzpicture}